\documentclass[longbibliography,twocolumn,amsmath,amssymb,amsfonts,superscriptaddress,floatfix,showpacs]{revtex4-1}
\usepackage[utf8]{inputenc}
\usepackage[english]{babel}
\usepackage[colorlinks=true,citecolor=green,linkcolor=red,urlcolor=blue]{hyperref}
\usepackage{braket}
\usepackage{bbm}
\usepackage{graphicx}
\usepackage{color}
\usepackage{dsfont}
\usepackage{float}
\usepackage{subcaption}
\newcommand{\skippart}[1]{}
\newcommand{\simplesection}[1]{\emph{#1} ---}

\usepackage{tikz}
\usetikzlibrary{er,positioning,arrows}

\begin{document}

\title{Cardy states, defect lines and chiral operators of coset CFTs on the lattice}
\author{Laurens~\surname{Lootens}}
\affiliation{Department of Physics and Astronomy, Ghent University, Krijgslaan 281, S9, B-9000 Ghent, Belgium}
\author{Robijn~\surname{Vanhove}}
\affiliation{Department of Physics and Astronomy, Ghent University, Krijgslaan 281, S9, B-9000 Ghent, Belgium}
\author{Frank~\surname{Verstraete}}
\affiliation{Department of Physics and Astronomy, Ghent University, Krijgslaan 281, S9, B-9000 Ghent, Belgium}

\begin{abstract}
We construct Cardy states, defect lines and chiral operators  for rational coset conformal field theories on the lattice. The bulk theory is obtained by taking the overlap between tensor network representations of different string-nets, while the primary fields emerge from using the topological superselection sectors of the anyons in the original topological theory. This mapping provides an explicit manifestation of the equivalence between conformal field theories in two dimensions and topological field theories in three dimensions: their groundstates and elementary excitations are represented by exactly the same tensors. 
\end{abstract}
\maketitle

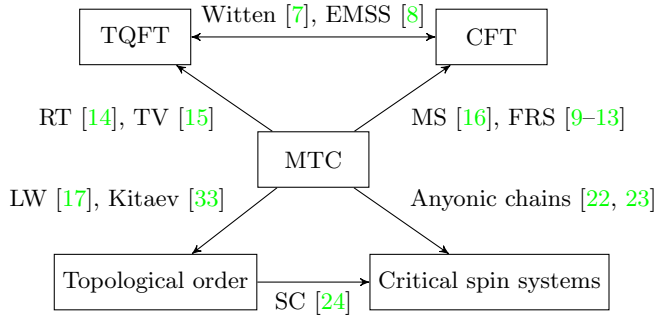
\begin{figure}
	\vspace{2pt}
	\begin{tikzpicture}[->,>=stealth',auto,node distance=1.5cm]
	\node[entity] (mtc) {MTC};
	\node[entity] (tqft) [above left=25pt and 25pt of mtc] {TQFT};
	\node[entity] (cft) [above right=25pt and 25pt of mtc] {CFT};
	\path (mtc) edge node [below left] {RT \cite{reshetikhin1991invariants}, TV \cite{turaev1992state}} (tqft);
	\path (mtc) edge node [below right] {MS \cite{moore1989classical}, FRS \cite{fuchs2002tft,fuchs2004tft1,fuchs2004tft2,fuchs2005tft,fjelstad2006tft}} (cft);
	\path [<->](tqft) edge node [above] {Witten \cite{witten1989quantum}, EMSS \cite{elitzur1989remarks}} (cft);
	\node[entity] (to) [below left=25pt and 0pt of mtc] {Topological order};
	\node[entity] (css) [below right=25pt and 0pt of mtc] {Critical spin systems};
	\path (mtc) edge node [above left] {LW \cite{levin2005string}, Kitaev \cite{kitaev2006anyons}} (to);
	\path (mtc) edge node {Anyonic chains \cite{feiguin2007interacting,buican2017anyonic}} (css);
	\path (to) edge node [below] {SC \cite{vanhove2018mapping}} (css);
	\end{tikzpicture}
	\caption{Schematic overview of the relations between the concepts described in the main text, in the continuum (top) and on the lattice (bottom).}
	\label{overview}
\end{figure}

\begin{figure}
\begin{subfigure}{0.35\linewidth}
\includegraphics[scale=1]{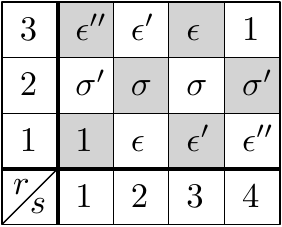}
\subcaption{}
\end{subfigure}%
\begin{subfigure}{0.6\linewidth}
\includegraphics[scale=1]{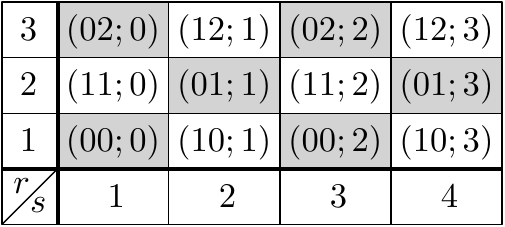}
\subcaption{}
\end{subfigure}
\caption{Kac table of the tricritical Ising model; conventional notation (a), coset notation (b).}
\label{kactricrit}
\end{figure}

\begin{figure}
\includegraphics{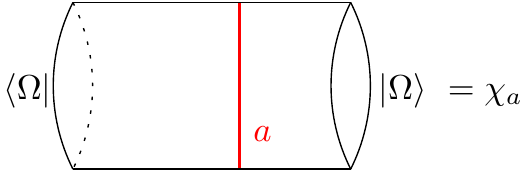}
\caption{Putting an MPO around the non-contractible loop of a cylinder with vacuum boundary conditions gives a single character partition function determined by the MPO, since pulling the MPO to the edge creates the boundary condition $\ket{a}$.}
\label{cylinder}
\end{figure}

\begin{figure}
	\begin{subfigure}[b]{0.5\linewidth}
		\includegraphics[scale=1]{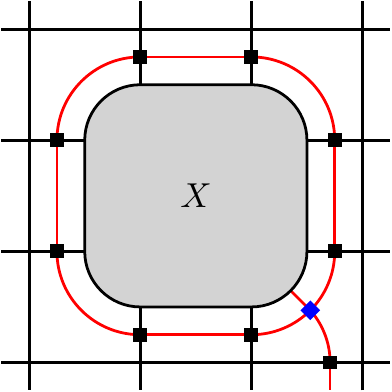}
		\subcaption{}
	\end{subfigure}%
	\begin{subfigure}[b]{0.3\linewidth}
		\includegraphics[scale=1]{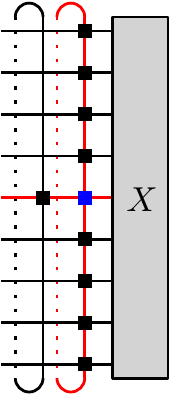}
		\subcaption{}
	\end{subfigure}
	\caption{Chiral operator insertion corresponding to the central idempotents on the lattice (a), eigenvector X of the transfer matrix with a defect projected onto one of the conformal towers (b). In the continuum, these two are related by a conformal transformation.}
	\label{primary1}
\end{figure}

\simplesection{Introduction}
In 1970, Kadanoff and Ceva \cite{kadanoff1971determination} constructed an operator algebra for the two-dimensional Ising model on the lattice which allowed for the determination of all critical indices. These results were later understood with the advent of conformal field theory (CFT) as being lattice representations of the continuum operators of the Ising CFT. Despite the vast body of literature on critical lattice systems \cite{fradkin2017disorder,fradkin1980disorder,mong2014parafermionic,hauru2016topological}, the general correspondence between lattice operators and their continuum counterparts has remained elusive. Part of the reason for this is the fact that CFT operators are in general chiral continuum fields, which by definition have to correspond to non-local operators on the lattice. Certain local fields in lattice Virasoro minimal models have been constructed by Pasquier \cite{pasquier1987operator} using what we now know as the modular $S$ matrix of a topological quantum field theory (TQFT), already hinting at a connection between CFT and TQFT. Shortly after Witten \cite{witten1989quantum} and subsequently Elitzur-Moore-Schimmer-Seiberg \cite{elitzur1989remarks} realised the strong connection between Chern-Simons theory on the one hand and Wess-Zumino-Witten (WZW) models on the other. Since then, the correspondence between TQFT and CFT has been the subject of thorough investigation culminating in a version of the holographic principle stating that all rational CFTs can be constructed holographically from a TQFT \cite{fuchs2002tft,fuchs2004tft1,fuchs2004tft2,fuchs2005tft,fjelstad2006tft}. This connection has proven to be extremely fruitful in the classification of rational CFTs and their topological properties such as defects, interfaces and boundaries; however, a general interpretation of this correspondence on the lattice is lacking.\\\\
The relevant mathematical structure in both TQFT and CFT is a modular tensor category (MTC). The Reshetekhin-Turaev/Turaev-Viro \cite{reshetikhin1991invariants,turaev1992state} constructions allow one to build a TQFT from any given MTC, while for a given CFT, the MTC is just a basis-invariant formulation of its Moore-Seiberg data \cite{moore1989classical,fuchs2002tft}. On the lattice, a large class of topological models can be constructed as Levin-Wen string net models \cite{levin2005string}, which take a tensor fusion category as input data and produce a topological phase described by the Drinfel'd double, which is a MTC. String-net models admit a very natural description in terms of tensor networks \cite{schuch2010peps,buerschaper2009explicit,csahinouglu2014characterizing,bridgeman2017anomalies}. The link between critical lattice models and MTC has been firmly established in the context of anyonic spin chains in (1+1)D \cite{feiguin2007interacting, buican2017anyonic}, in which criticality was shown to be protected by the presence of topological symmetries. In Figure \ref{overview}, a schematic overview of these relations is given. Recent progress was made towards further establishing these correspondences by use of the concept of a ``strange correlator'' \cite{vanhove2018mapping}, first developed for symmetry protected topological phase detection in \cite{you2014wave}. The strange correlator maps a (2+1)D topological projected entangled pair state (PEPS) wave function to a 2D critical lattice model. This is done by taking the overlap of the topological PEPS with a trivial product state, which results in a partition function of a 2D lattice model that is either critical or symmetry broken \cite{vanhove2018mapping}. In this work, we extend this procedure by considering the string-net PEPS description of the interface between two topological phases and show that it corresponds to the coset construction of CFTs, one of the main model building tools for rational CFTs. The full set of conformal defects, conformal boundary conditions and chiral operators of the critical lattice model are shown to correspond to the virtual matrix product operator (MPO) symmetries and anyonic excitations characterising the two topological PEPS wave functions \cite{bultinck2017anyons, williamson2017symmetry}. To illustrate our results, we turn to the minimal models and study a well known lattice representation known as the restricted solid-on-solid (RSOS) models \cite{andrews1984eight, baxter1982inversion}. We shed new light on the known conformal boundary conditions of these theories \cite{behrend2000boundary, cardy1989boundary} and write down a general form for the chiral operators. Similar results have been announced by Aasen, Mong and Fendley in \cite{aasen2016topological} by characterizing defect lines in height models. It is not clear how the coset construction, discussed in this paper, fits into their formalism.\\
\simplesection{Coset CFT}
Cosets allow for the construction of a large class of rational conformal field theories, starting from two WZW models $G_k$ and $H_k$ based on Lie groups $G$ and $H$ with $H \subset G$. The coset $G_k/H_{k'}$ then corresponds to the embedding $H_{k'} \subset G_k$, which can be interpreted as gauging the $H$ subgroup of $G$ in the WZW model based on $G$. In \cite{bais2009condensate} it was shown that the coset construction can be described in terms of quantum group symmetry-breaking in the tensor product theory $G_k \otimes \overline{H_{k'}}$. The spectrum, fusion rules and modular properties of fields in the $G_k/H_{k'}$ coset can be obtained through the condensation of bosonic simple currents in the tensor product theory. In particular, the \textit{field identifications} of the coset CFT are such that fields related by fusion with any of the condensed bosons are identified, while the \textit{branching rules} are enforced by requiring that any field that has a non-trivial monodromy with any of the condensed bosons is confined. If we now consider the ground states of topologically ordered systems $\ket{\psi_G}$ and $\ket{\psi_H}$ described by the representation theory of the quantum groups $G_k$ and $H_{k'}$ respectively, we can take the overlap of these two states provided we insert some local isometry $V$ to account for the mismatch in Hilbert spaces. This can be written as
\begin{equation}
\bra{\psi_H}V\ket{\psi_G} = \bra{\Omega}(\ket{\psi_G}\otimes\ket{\overline{\psi_H}}),
\end{equation}
which corresponds to taking the overlap of some product state with a topological state described by the tensor product theory $G_k \otimes \overline{H_{k'}}$. This is precisely the strange correlator method for obtaining critical lattice systems from topologically ordered states, and the associated CFT is expected to corespond to the WZW coset $G_k/H_{k'}$. The condensation and confinement mechanisms are implemented by the specific choice of product state; in essence, the degrees of freedom in the resulting partition function are restricted to the fields allowed in the coset CFT.\\
\simplesection{Virasoro minimal models} To illustrate this procedure, we move to the Virasoro minimal models; these are in many aspects the simplest class of CFTs with a finite number of representations of the Virasoro algebra. The most well known among these is the Ising CFT, denoted $\mathcal{M}(3,4)$, and in general the minimal models are written as $\mathcal{M}(p,q)$ with central charge $c = 1 - 6(p-q)^2/pq$,
with $p$ and $q$ two coprime integers larger than $1$. The representations of these CFTs arrange themselves in the \textit{Kac table} and they are labeled by two integers $r \in [1,q-1]$ and $s \in [1,p-1]$, with conformal dimensions given by $h_{r,s} = ((pr - qs)^2 - (p-q)^2)/4pq$. We will restrict to the unitary minimal models, for which $|p-q| = 1$, and write $p = k+2$ and $q = k+3$. In terms of the coset construction, these models can be written as
\begin{equation}
\mathcal{M}(k+2,k+3) = \frac{\text{su}(2)_1 \otimes \text{su}(2)_k}{\text{su}(2)_{k+1}}.
\end{equation}
Following \cite{bais2009condensate}, we will denote the $2(k+1)(k+2)$ fields of this coset as $(ab;c) \in \text{su}(2)_1 \otimes \text{su}(2)_k \otimes \overline{\text{su}(2)}_{k+1}$. In general, this construction yields one non-trivial boson $(1 k; (k+1))$ that we can condense. Defining a parity $(a + b + c) \mod 2$, after condensation of this boson all the fields with odd parity are confined, while the field identification yields the Kac table symmetry $h_{r,s} = h_{q-r,p-s}$. This results in $(k+1)(k+2)$ pairwise identified fields that form a representation of $\text{su}(2)_k \otimes \text{su}(2)_{k+1}$, which is indeed the chiral algebra of the Virasoro minimal models.\\\\
To interpret this construction using the strange correlator, we start with a topological string-net PEPS wave function using $\text{su}(2)_1 \otimes \text{su}(2)_k \otimes \overline{\text{su}(2)}_{k+1}$ as input tensor fusion category. The product state is chosen in the even parity subsector, which means the resulting partition function becomes a direct sum over the even and odd parity loops. If we now fix these loops to only be in the even sector, corresponding to the notion of symmetry enriched topological order (SET) \cite{williamson2017symmetry}, we break precisely those MPO symmetries corresponding to the confined anyons since these would change the parity of the loops. Besides these obvious choices, the only remaining freedom in the product state is determined by the rules of the desired critical lattice model. The models we will map to are the RSOS models, which are known to correspond to minimal models in their continuum limit. These models describe a square lattice of fluctuating height variables that take on a discrete set of values, with the restriction that neighbouring lattice sites can only differ by one. Although our approach works for all RSOS models, we will illustrate with the simplest non-trivial example and state the generalities afterwards.\\
\simplesection{Ising and tricritical Ising model}
The first non-trivial minimal model is the Ising model, which was already described in earlier work without the coset construction. The reason that the Ising model does not require this extra machinery is that it admits a different representation as the following coset \cite{francesco2012conformal}:
\begin{equation}
\mathcal{M}(3,4) = \frac{(E_8)_1 \otimes (E_8)_1}{(E_8)_2},
\end{equation}
with $E_8$ the exceptional Lie group of rank 8. This coset is rather trivial, since the WZW model associated to $E_8$ at level $1$ is a $c = 8$ CFT with only the identity representation, which is a trival theory. This means that the coset construction reduces to $(E_8)_2$, which has representations that fuse according to the Ising fusion algebra. This is also the case for $\text{su}(2)_2$, but in contrast, $(E_8)_2$ does have the correct Frobenius-Schur indicator so it suffices to consider $(E_8)_2$ by itself, which is what was done in \cite{vanhove2018mapping}. We therefore turn to the tricritical Ising model, $\mathcal{M}(4,5)$, which has the Kac table shown in Figure \ref{kactricrit}. In the coset construction, the $(12;3)$ anyon is a boson that after condensation is identified with the vacuum.
An example of a critical model in the tricritical ising universality class is the $A_4$ RSOS model, which can be written as a strange correlator using the following product state on the string-net PEPS:
\begin{equation}
W\begin{pmatrix}
a & b\\
d & c
\end{pmatrix} =
\begin{gathered}
\includegraphics{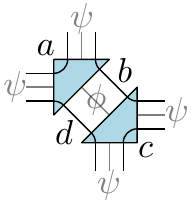}
\end{gathered},
\end{equation}
As mentioned above, adjacent heights are required to differ only by one, which is imposed by putting $\psi = \ket{10;1}$. The product state for $\phi$ can then be determined by requiring this model to be isotropic, or equivalently invariant under rotations over 90 degrees. The rotated tensors can be related to each other via $F$- moves, and requiring equality gives $\phi = (1 +\frac{1+\sqrt{5}}{2}) \ket{00;0} + \ket{00;2}$. Since this model is a 4-level system, we additionally restrict the loops $\{a,b,c,d\}$ to the bottom row of the Kac table. This tensor then coincides with the standard Boltzmann weight for the $A_4$ RSOS model, but only up to minus signs. These minus signs drop out when contracting the tensor network to obtain the partition function since they always come in pairs, but they are crucial because without them the partition function does not have MPO symmetries. The restriction to the bottom row of the Kac table has the consequence that all MPO symmetries with $r \neq 1$ are broken in the bulk. We expect that such MPO symmetries should exist in this model, but they are not given in a simple form in this formalism. It is important however to realise that this is purely a consequence of considering the RSOS models; in general, we expect that critical lattice models based on the coset strange correlator have all the required MPO symmetries. Proceeding with the MPO symmetries with $r=1$ for the $A_4$ RSOS model, we can calculate twisted conformal spectra and project onto the different conformal towers using the central idempotents of the tube algebra. The lack of $r \neq 1$ MPOs means we cannot resolve the different $r$ sectors in this model.
\\
\simplesection{Cardy states}
So far we have only looked at critical partition functions on a square lattice with periodic boundary conditions in both directions, i.e. on a manifold with the topology of a torus. Many interesting physical systems however do not have this topology, and it is more natural to define them on a cylinder with boundary conditions on the ends. The study of boundary conformal field theory (BCFT) was initiated by Cardy in 1989 \cite{cardy1989boundary} by looking at the constraints imposed by conformal invariance, and gained significance because it revealed the algebraic structure of CFT in a more direct manner. The resulting conformal boundary conditions, also called \textit{Cardy states}, gain a very natural interpretation in the strange correlator language. Assuming we have mapped a topological string-net PEPS on a cylinder to a critical partition function, boundary conditions can in general be described by a matrix product state (MPS) with periodic boundary conditions on the edges of the cylinder. These boundary conditions should be RG fixed points, and it was shown in \cite{verstraete2005renormalization} that any MPS will flow to a product state, which means that all these boundary conditions have to be product states. We can now define the vacuum as the boundary condition that yields all the other boundary conditions when acted upon by an MPO along a non-contractible loop of the cylinder, as shown in Figure \ref{cylinder}. Given that all the boundary conditions must be product states, this uniquely fixes the the vacuum up to group automorphisms of the fusion algebra. For the RSOS models, it was shown that the conformal boundary conditions coincide with the integrable boundary conditions, that satisfy a boundary version of the Yang-Baxter equation \cite{behrend1998classification}. Applying our procedure for these models, we recover the same integrable boundary conditions without invoking the integrability of the model. To illustrate, for the example of the $A_4$ RSOS model, we define the vacuum boundary condition on $N$ sites as the product state
\begin{multline}
\ket{\Omega} = \Big(\ket{00;0}+\ket{10;1}+\ket{01;1}\\
+\ket{11;2}+\ket{02;2}+\ket{12;3}\Big)^{\otimes N}.
\end{multline}
The new boundary conditions are then generated by acting on this with the different MPOs, given that we fix the bulk loops to be in the bottom row of the Kac table. We expect that the terms in the vacuum from higher rows of the Kac table are only necessary because some MPO symmetries are broken in the bulk, and we expect these will no longer be necessary once we find an explicit form for all MPO symmetries in this RSOS model. For general critical models, all MPOs will be present in the bulk and this complication will not arise.\\
\simplesection{Orbifolding}
Besides the coset construction, other well known concepts in CFT can also be understood in terms of gauging of quantum group symmetries \cite{bais2009condensate}. Given some RCFT with a discrete symmetry group $\mathcal{G}$, the \textit{orbifold} CFT is constructed by modding out $\mathcal{G}$ and projecting the Hilbert space onto the $\mathcal{G}$ invariant subspace. Considering as an example the tetracritical Ising model $\mathcal{M}(5,6)$, there is a discrete $\mathds{Z}_2$ symmetry generated by fusion with the simple current $(00;4)$. This is a bosonic representation which can therefore be condensed, after which we should require that the partition function is restricted to the $\mathds{Z}_2$ invariant subspace. This can be done by the insertion of the following projector in the partition function:
\begin{equation}
\mathcal{O}_A = \frac{1}{2}\left(A_{0000} + A_{0404} + A_{4044} + A_{4440}\right)
\label{projector}
\end{equation}
where we used the shorthand notation $(00;0) \equiv 0$ and $(00;4) \equiv 4$ and the tubes $A_{abcd}$ as defined in \cite{bultinck2017anyons}. This corresponds to the insertion of the projector $P = (1/2)(O_0 + O_4)$ along the two non-contractible loops of the torus, and is equivalent to the orbifold construction of the three state Potts model in \cite{affleck1998boundary} as a sum of twisted partition functions. We note that this construction coincides with the simple current extension discussed in \cite{fuchs2002tft,fuchs2004tft1,fuchs2004tft2,fuchs2005tft,fjelstad2006tft}.
\\
\simplesection{Chiral operators} As mentioned above, the central idempotents $\mathcal{P}_{a\bar{b}}$ of the tube algebra can be used to project onto different conformal towers analogously to how they project onto the different topological sectors in the topological PEPS. In \cite{bultinck2017anyons} it was shown that the same idempotents can be used to write down a general form for the anyonic excitations, by placing the corresponding central idempotent around some of the PEPS tensors. These anyonic excitations have a certain topological spin, and they change non-trivially when they are pulled through an MPO symmetry. Importantly, these calculations can be done exactly in the MPO formalism. These concepts have a very close counterpart in conformal field theory, where it is also known that primary operators have a non-trivial conformal spin and these operators change non-trivially when they are pulled through a topological defect. It is therefore very natural to assume that, after applying the strange correlator mapping, the anyonic excitations in the topological PEPS will map to primary operators in the CFT. The only question that remains then is which specific tensor $X$ to insert into the anyonic excitation tensor in Figure \ref{primary1}. To answer this, we notice that this situation is analogous to the case where we project the transfer matrix onto one of the conformal towers using a central idempotent. To some extent, this reasoning can be seen as a lattice version of the conformal mapping of a punctured plane to a cylinder. We expect that if we take the tensor $X$ to be one of the eigenvectors of the product of the transfer matrix with the central idempotent, we can write down a general form for any operator in the conformal tower. This eigenvector will only be exact in the thermodynamic limit of a transfer matrix defined on a ring of $N \rightarrow \infty$ sites, and for a given size $N$ the approximation will be worse for larger conformal weights. This in turn implies that to get an accurate representation of the operators in the conformal tower, we will have to grow the idempotent to include more sites as we go up the conformal tower. This implies that on the lattice, primary operators with a larger conformal weight will have a larger support, while in the continuum field theory limit all these operators are pointlike.\\
A particularly important primary operator is the energy-momentum tensor, with holomorphic and antiholomorphic components given by the second descendants of the vacuum primary. The components of this tensor can therefore also be written in terms of a central idempotent as a state in the vacuum tower. This object has also been written down by Kadanoff-Ceva for the Ising model \cite{kadanoff1971determination} and by Koo-Saleur for RSOS models \cite{koo1994representations, milsted2017extraction}, and it would be interesting to interpret their results in this language.\\ 

\simplesection{Conclusion and outlook} We have constructed Cardy states, topological defects and chiral operators of rational coset CFTs on the lattice. Using a generalised strange correlator as the overlap between two states in different topological phases we map to a critical lattice model exhibiting non-local MPO symmetries. Using these MPOs we can construct boundary conditions and chiral operators using the central idempotents of the tube algebra. We illustrated this procedure for the tricritical Ising model, where our particular choice of lattice model breaks several of these MPO symmetries such that we do not recover the full set of topological defects for this CFT, nevertheless the full set of boundary conditions is recovered. We plan to discuss how these missing topological defects could be recovered by making use of the concept of bimodules in future work \cite{kitaev2012models}. Future work will tell if these missing defects admit a simple lattice representation in this formalism, but it is important to note that for a generic critical model built using this approach, we will recover all topological defects. While the present work is mainly conceptual in nature, the advantage of having lattice realisations of these CFT concepts allows for explicit numerical computations for which the tensor network toolbox is perfectly suited.

\begin{acknowledgments}
We thank Paul Fendley, J\"urgen Fuchs, Christophe Schweigert, Andreas Ludwig, Dominic Williamson, Nick Bultinck, Markus Hauru and Jutho Haegeman for insightful discussions. This work is supported by an Odysseus grant from the FWO, ERC grant QUTE (647905), BeyondC Special Research Programme, FWO project (G087918N)  and FWO PhD-grant (R.V).
\end{acknowledgments}
\bibliographystyle{apsrev4-1}
\bibliography{main.bib}
\end{document}